\documentclass[twoside]{article}
\usepackage{fleqn,espcrc2}

\newcommand{\msbar}{$\overline{\mbox{MS}}$ }
\newcommand{\be}{\begin{equation}}
\newcommand{\ee}{\end{equation}}

\newcommand{\bmG}{\mathbf G}

\begin{document}

\title{ Bottom Quark Mass Determination from low-$n$ Sum Rules}

\author{ {G. Corcella and A.H. Hoang} 
\address{ Max-Planck-Institut f\"ur Physik,
Werner-Heisenberg-Institut, 
\\ F\"ohringer Ring 6, D-80805 M\"unchen, Germany }}

\begin{abstract}
We study the uncertainties in the \msbar bottom quark mass determination 
using relativistic sum rules to ${\cal O}(\alpha_S^2)$. We include charm mass
effects and secondary $b\bar b$ production and treat the
experimental continuum region more conservatively than
previous analyses. The PDG treatment of the
region between the resonances $\Upsilon (4S)$ and $\Upsilon (5S)$ is 
reconsidered. Our final result reads: 
$\bar m_b(\bar m_b)=(4.20\pm 0.09)~\mathrm{GeV}$.
\vskip-95mm
   {\small \noindent
  Talk given at QCD 03, Montpellier, France, 2-9 July 2003.}
   \begin{flushright} {\large MPP-2003-119} \\
  \end{flushright}
\vskip75mm 
\end{abstract}

\maketitle

For the sake of reliable measurements at the current $B$-factory
experiments, 
a precise knowledge of the bottom quark mass $m_b$ will be essential.
In particular, the precision on 
the extraction of the Cabibbo--Kobayashi--Maskawa matrix 
elements $V_{ib}$ from the data will depend on the uncertainty
on the bottom mass. 
For example, an error of 60 MeV in $m_b$ leads to a $3\%$ uncertainty 
in $V_{ub}$ from the semileptonic partial width 
$\Gamma(B\to X_u\ell\nu)$ \cite{batta}.

In this talk we present the results of a detailed compilation of
uncertainties in the \msbar bottom quark mass \cite{corh}. Our analysis
is more conservative than an earlier one in
Ref.~\cite{ks} and includes a number of effects that
were previously neglected. 
Our method consists of 
determining the $b$ mass by fitting the experimental moments
of the $b\bar b$ cross section in $e^+e^-$ annihilation
to their correspondent theoretical expressions.
The moments are defined as follows:
\be
P_n =  \int\frac{ds}{s^{n+1}}\,R_{b\bar b}(s),
\ee
where $R_{b\bar b}=\sigma(e^+e^-\to b\bar b+X)/\sigma(e^+e^-\to\mu^+\mu^-)$.
The virtual $Z$ contribution is strongly suppressed and neglected.

We use low-$n$ ("relativistic") moments, i.e. $n\leq 4$.  
Relativistic moments exhibit the nice feature that they are 
dominated by scales of order $m_b$ and that they can be computed in
fixed-order perturbation theory.  
However, they have the disadvantage that 
they strongly depend on the badly known experimental continuum region.
As for the $b$-mass definition, we adopt the \msbar mass, since it is an 
appropriate definition for processes where $b$ quarks are off-shell.

A compilation of the theoretical moments up 
to ${\cal O}(\alpha_S^2)$ was given in \cite{ks}.
In our work \cite{corh} we also included the effects at ${\cal O}(\alpha_S^2)$
of the non-zero charm mass and of secondary
$b\bar b$ production, with the $b\bar b$ pair coming from gluon radiation
off light quarks.
Referring, for semplicity, to $\bar m_b(\bar m_b)$ the 
theoretical moments have a simple form:
\begin{eqnarray}
P_n 
&=&
\frac{1}{(4\bar m_b(\bar m_b))^{n}}\bigg\{f_n^0 +  
\bigg(\frac{\alpha_s(\mu)}{\pi}\bigg)f_n^{10} 
\nonumber\\[2mm]
&+& 
\bigg(\frac{\alpha_s(\mu)}{\pi}\bigg)^2
  \bigg(f_n^{20}(r)-  
         \frac{1}{4}\beta_0\,f_n^{10}
   \ln\Big(\frac{\bar m_b^2(\bar m_b)}{\mu^2}\Big)
  \bigg)
\nonumber\\[2mm]
&+&
\frac{\langle{\textstyle\frac{\alpha_s}{\pi}}\bmG^2\rangle}
{(4\bar m_b(\bar m_b))^2}
  \bigg[g_n^{0} + 
\bigg(\frac{\alpha_s(\mu)}{\pi}\bigg) g_n^{10} 
  \bigg]\bigg\}.
\label{mom}
\end{eqnarray} 
In Eq.~(\ref{mom}), $r=m_c/m_b$, 
$\mu$ is the renormalization scale and we have 
included the contribution from the dimension four gluon 
condensate \cite{cond}. The general expression of $P_n$ for
$\bar m_b(\mu)$ can be found in Refs.~\cite{corh,ks}.
Results for the coefficients $f_n$ are reported in Table~\ref{tmom}.
We point out that the contribution of secondary $b\bar b$ production and
charm mass affects only $f_n^{20}$.
Such effects turn out to be small, as can be seen comparing the 
values for $f_n^{20}$ in Table~\ref{tmom} with the ones of
Ref.~\cite{ks}.
\begin{table*}[ht]
\small
\caption{Coefficients of the theoretical moments in Eq.~(\ref{mom}).} 
\vskip 3mm
\begin{center}
\begin{tabular}{|c||c|c|c|c|} \hline
$n$ & $1$&$2$&$3$&$4$\\ \hline\hline 
$f_n^{0}$&$0.2667$&$0.1143$&$0.0677$&$0.0462$ \\ \hline 
 $f_n^{10}$& $ 0.6387$&$ 0.2774$&$ 0.1298$&$ 0.0508$ \\ \hline 
 $f_n^{11}$&$ 0.5333$&$ 0.4571$&$ 0.4063$&$ 0.3694$ \\ \hline 
 $f_n^{20}(0)$&$ 0.9446$&$ 0.8113$&$ 0.5172$&$ 0.3052$ \\ \hline 
 $f_n^{21}$&$ 0.8606$&$ 1.2700$&$ 1.1450$&$ 0.8682$ \\ \hline 
 $f_n^{22}$&$ 0.0222$&$ 0.4762$&$ 0.8296$&$ 1.1240$ \\ \hline 
\end{tabular}\label{tmom}
\end{center}
\end{table*}
Table~\ref{tab1} displays the impact of charm mass corrections in terms of
$\Delta f_n=f_n^{20}(r)-f_n^{20}(0)$. The smallness of
$c$-mass effects is however strongly related to the use of the \msbar
mass definition which we have adopted.
In fact, if we had chosen the pole scheme for the bottom mass, the 
inclusion of the charm mass would have had a much stronger impact, as shown in
Table~\ref{tab2}. 
This can be understood from the fact that the finite charm mass
represents an infrared cut-off in the loop integrations and that the
pole-mass definition is much more sensitive to infrared momenta.
\begin{table*}[ht]  
\small
\caption{
Corrections due to the non-zero charm quark mass in the \msbar scheme.}
\vskip 3mm
\begin{center}
\begin{tabular}{|c||c|c|c|c|c|} \hline
$r$ & $0.1$ & $0.2$ & $0.3$ & $0.4$ & $0.5$\\ \hline\hline
 $\Delta f_1$ & $-0.0021$ & $-0.0078$ & $-0.0164$ &
 $-0.0266$ & $-0.0382$ 
\\ \hline 
 $\Delta f_2$ & $-0.0028$ & $-0.0091$ & $-0.0187$ &
 $-0.0302$ & $-0.0430$ 
\\ \hline 
 $\Delta f_3$ & $-0.0024$ & $-0.0101$ & $-0.0204$ &
 $-0.0330$ & $-0.0466$ 
\\ \hline 
 $\Delta f_4$ & $-0.0030$ & $-0.0109$ & $-0.0219$ &
 $-0.0348$ & $-0.0491$ 
\\ \hline
\end{tabular}\label{tab1}
\end{center}
\end{table*}
\begin{table*}[ht] 
\small
\caption{
As in Table~\ref{tab1}, but in the pole-mass scheme.}
\vskip 3mm
\begin{center}
\begin{tabular}{|c||c|c|c|c|c|} \hline
$r$ & $0.1$ & $0.2$ & $0.3$ & $0.4$ & $0.5$\\ \hline\hline
$\Delta f_1$ & $0.0809$ & $0.1505$ & $0.2113$ &
$0.2656$ & $0.3145$ 
\\ \hline 
$\Delta f_2$ & $0.0684$ & $0.1267$ & $0.1765$ &
$0.2203$ & $0.2593$ \\ \hline 
$\Delta f_3$ & $0.0608$ & $0.1106$ & $0.1531$ &
$0.1896$ & $0.2221$ \\ \hline 
$\Delta f_4$ & $0.0545$ & $0.0988$ & $0.1358$ &
$0.1676$ & $0.1764$ 
\\ \hline
\end{tabular}\label{tab2}
\end{center}
\end{table*}
\begin{table*}\small
\caption{Individual 
contributions to the experimental moments including uncertainties.
In the continuum the displayed uncertainties are the theoretical ones only.}
\vskip 3mm
\begin{center}
\begin{tabular}{|c||c|c|c|c|} \hline
 &$P_1$&$P_2$&$P_3$&$P_4$\\
\raisebox{1.5ex}[-1.5ex]{contribution} & 
 $\times\,10^3~\mbox{GeV}^2$ & $\times\,10^5~\mbox{GeV}^4$ & 
 $\times\,10^7~\mbox{GeV}^6$ & $\times\,10^9~\mbox{GeV}^8$ \\
\hline\hline 
$\Upsilon(1S)$ & $ 0.766(29)$ & $ 0.856(32)$ & $ 0.956(36)
$ & $ 1.068(40)$ 
\\ \hline 
$\Upsilon(2S)$ & $ 0.254(16)$ & $ 0.252(16)$ & 
$ 0.251(15)$ & $ 0.250(15)$ 
\\ \hline 
$\Upsilon(3S)$ & $ 0.211(29)$ & $ 0.196(27)$ & 
$ 0.183(26)$ & $ 0.171(24)$ 
\\ \hline 
$[\Upsilon(\mbox{4S})-\Upsilon(\mbox{5S})]$ & 
$ 0.251(95)$ & $ 0.218(82)$ & $ 0.190(72)$ & $ 0.165(62)$ 
\\ \hline 
$\Upsilon(\mbox{6S})$ & $ 0.048(11)$ & $ 0.039(9)$ & $ 0.032(7)$ & $ 0.027(6)$ 
\\ \hline 
$11.1~\mbox{GeV}-12.0~\mbox{GeV}$ & $ 0.418(57)$ & $ 0.314(44)$ & 
$ 0.236(34)$ & $ 0.178(27)$ 
\\ \hline 
$12.0~\mbox{GeV}-M_Z$ & $ 2.467(26)$ & $ 0.886(21)$ 
& $ 0.414(13)$ & $ 0.217(8)$ 
\\ \hline 
$M_Z-\infty$ & $ 0.047(1)$ & $ 0.000(0)$ & $0.000(0)$ & $0.000(0)$ 
\\ \hline
\end{tabular}\label{tab3}
\end{center}
\end{table*}
\begin{table*}
\small
\caption{Central values and uncertainties for $\bar m_b(\bar m_b)$.}
\vskip 3mm
\begin{center}\label{tab4}
\begin{tabular}{|c||c|c|c||c|c|} \hline
 &\multicolumn{3}{|c||}{Method 1 (2)}& 
   \multicolumn{2}{|c|}{Method 3 (4)}
\\ \hline
$n$&1&2&3&1 & 2 
\\ \hline\hline
central&4210(4214)&4200(4205)&4197(4200)&4191(4195)&4191(4191)
\\ \hline\hline
$\Upsilon(1S)$&14 (13)&12 (12)&11 (11)&11 (11)&9 (9)
\\ \hline
$\Upsilon(2S)$&7 (7)&6 (6)&5 (5)&4 (4)&3 (3) 
\\ \hline
$\Upsilon(3S)$&14(14)&10 (10)&8 (8)&7 (7)&3 (3) 
\\ \hline
$4S-5S$&45 (44)&32 (32)&22 (22)&18 (18)&4 (4) 
\\ \hline
$\Upsilon(6S)$&5 (5)&3 (3)&2 (2)&2 (2)&0 (0)
\\ \hline\hline
combined&67 (67)&50 (50)&38 (38)&33 (33)&15 (15)
\\ \hline\hline
[region 1]$_{\rm th}$&27$_{\rm th}$ 
(26$_{\rm th}$)&17$_{\rm th}$ (17$_{\rm th}$)&11$_{\rm th}$ 
(11$_{\rm th}$)&7$_{\rm th}$ (7$_{\rm th}$) & 2$_{\rm th}$ (2$_{\rm th}$)
\\ \hline
[region 2]$_{\rm th}$&12$_{\rm th}$(12$_{\rm th}$)&8$_{\rm th}$ 
(8$_{\rm th}$) &4$_{\rm th}$ (4$_{\rm th}$)&4$_{\rm th}$ 
(4$_{\rm th}$)&4$_{\rm th}$ (4$_{\rm th}$)
\\ \hline
[region 2]$_{10\%}$&115 (114)&33 (33)&13 (13)&49 (49)&29 (29)
\\ \hline 
[region 3]$_{\rm th}$&1$_{\rm th}$ (1$_{\rm th}$)&0$_{\rm th}$ 
(0$_{\rm th}$)&0$_{\rm th}$ 
(0$_{\rm th}$)&1$_{\rm th}$ (1$_{\rm th}$)&0$_{\rm th}$ (0$_{\rm th}$)
\\ \hline
[region 3]$_{10\%}$&2 (2)&0 (0)&0 (0)&2 (2)&0 (0)
\\ \hline
$\delta m_c$&0 (0)&0 (0)&0 (0)&0 (0)&0 (0)
\\ \hline
$\delta\alpha_s(M_Z)$&17 (18)&10 (11)&6 (6)&3 (3)&2 (2)
\\ \hline
$\delta\mu$&23 (5)&16 (14)&11 (27)&15 (27)&3 (50)
\\ \hline\hline
combined &184 (166)&77 (75)&41 (57)&76 (88)&37 (85)
\\ \hline\hline
total&251 (233)&127 (125)&79 (95)&110 (121)&51 (99) 
\\ \hline
\end{tabular}
\end{center}
\end{table*}
To evaluate the experimental moments, we consider 
the region of the resonances $\Upsilon (1S)-\Upsilon (6S)$ and the continuum.
We compute the moments of a generic resonance $k$ in the narrow width
approximation, i.e.
\be 
(P_n)_k={{9\pi\Gamma_k^{e^+e^-}}\over{\alpha(10~\mathrm{GeV})m_k^{2n+1}}},
\ee
where $\Gamma_k^{e^+e^-}$ is the partial $e^+e^-$ width for the
$k$-th resonance.
For the $\Upsilon(1S)$, $\Upsilon (2S)$, $\Upsilon(3S)$ and 
$\Upsilon(6S)$ we use the averages for masses and widths quoted in the
PDG \cite{pdg}. 
For the region between the 
$\Upsilon(4S)$ and the $\Upsilon(5S)$, i.e. between 10.5 and 10.95 GeV, 
we do not use the PDG averages, which were based on
results from CUSB \cite{cusb} and CLEO \cite{cleo} Collaborations. Both
experiments observed an enhancement at about 10.7 GeV. While CUSB did
not assign the enhancement to any resonance, CLEO fitted it to
a "$B^*$" resonance with mass
$m_{B^*}=10.684\pm 0.013$~GeV and $e^+e^-$ width 
$\Gamma^{e^+e^-}_{B^*}= 0.20\pm 0.11$~keV.
The PDG averages, on the other hand, ignore the $B^*$ results and,
therefore, lead to a contribution to $P_n$ that is smaller
than the original CUSB and CLEO data. In our analysis we took the
averages from the original CUSB and CLEO data, assuming the larger
uncertainties from CLEO (See Ref.~\cite{corh} for more details).

As far as the continuum is concerned, we subdivide it into three parts: 
11.1-12.0 GeV (region 1), where possible data may come from CLEO;
12 GeV - $M_Z$ (region 2) and above $M_Z$ (region 3).
There is no direct experimental data in the region above 11.1 GeV.
Nevertheless the measurements of $R_b$ by LEP I and LEP II agree with the
perturbative QCD prediction within $1\%$ at $M_Z$ and $10\%$ in the region
between 133 and 207 GeV explored by LEP II.
It is therefore not unreasonable to rely on perturbative QCD 
to estimate the contribution to the experimental moments above the
$\Upsilon(6S)$. In the analysis of Ref.~\cite{ks} the small
theoretical errors in the continuum region were inherently taken as the
experimental uncertainties. Since this leads to an implicit 
model-dependence, we adopt a more transparent treatment and take an
assigned fraction of the theoretical prediction as the experimental
uncertainty of the continuum. In this way the impact of the unknown 
experimental continuum contribution can be traced more easily.
The experimental moments are quoted in Table~\ref{tab3}. 
The uncertainties displayed for the three continuum regions 
are the ones from the theoretical uncertainties only.

For the determination of $\bar m_b(\bar m_b)$ and 
of the corresponding uncertainties,
we use four methods: we fit single moments and get directly 
$\bar m_b(\bar m_b)$ (method 1); we determine $\bar m_b(\mu)$ from
single-moment fits and subsequently $\bar m_b(\bar m_b)$ using renormalization
group equations (method 2); we fit the ratio $P_n/P_{n+1}$ and get 
$\bar m_b(\bar m_b)$ (method 3); we determine $\bar m_b(\mu)$ by fitting
$P_n/P_{n+1}$ and compute $\bar m_b(\bar m_b)$ using renormalization group 
equations (method 4).
We employ four-loop renormalization group equations, vary the renormalization
scale $\mu$ between 2.5 and 10 GeV and use 
$\alpha_s(M_Z)=0.118\pm 0.003$,
$m_c=1.3\pm 0.2$~GeV, 
$\langle{\textstyle\frac{\alpha_s}{\pi}}\bmG^2\rangle
=(0.024\pm 0.024)~\mbox{GeV}^4$ as theoretical input.
The results of our analysis are shown in Table~\ref{tab4}. The central values 
for $\bar m_b(\bar m_b)$ according to the four methods
agree within 15 MeV. 
For the errors coming from the experimental-continuum regions 2 and 3
we quote both, the one coming from the theory uncertainties shown in
Table~\ref{tab3} and  the error corresponding to a $10\%$ variation of the
theoretical prediction. The latter error scales roughly linearly,
i.e. assuming a $5\%$ ($20\%$) fraction decreases (increases) the error by a
factor of two.
In order to get the combined errors, in
the resonance region we treat half of the errors as uncorrelated
(added linearly) and half of the errors as correlated (added quadratically).
The errors in the continuum do not have any statistical correlation, hence
we add them linearly. Moreover, we add linearly 
the errors coming from the resonance and from the continuum regions.

We note that the errors yielded by fits of the first two moments $P_1$
and $P_2$ are rather large. As for the results given by fits of the 
moment ratios, the fit of $P_2/P_3$ using method 3
yields a rather small error of about 50 MeV. 
However, this result holds only if the same value of $\mu$ is chosen for 
both $P_2$ and $P_3$; 
a larger error would instead be found using independent values
of $\mu$ for the numerator and denominator of the ratio.
Since we believe that $P_3$ can be calculated reliably using
both methods 1 and 2, we adopt the error on $P_3$ as our final estimate of the
uncertainty in the \msbar bottom mass determination. Rounding to units 
of 10 MeV, we obtain:
\be
\bar m_b(\bar m_b)=(4.20\pm 0.09)~\mathrm{GeV},
\ee
assuming a $10\%$ error for the experimental continuum regions 2 and 3. 
Within the error range, our result is in agreement with the estimate
of \cite{ks}. Our error is nonetheless larger than the 50 MeV of
Ref.~\cite{ks}, which is due to the 
different treatment of the resonance region and to the more conservative
choice for the experimental error in the continuum region.

\end{document}